\documentclass[conference,A4paper]{IEEEtran}

\usepackage{graphicx,subfigure}
\usepackage{amsfonts,amsmath,amssymb, mathrsfs}
\usepackage{multirow}
\usepackage{epic,eepic,eepicemu}
\usepackage{epsf}
\usepackage{epsfig}
\usepackage{graphics,subfigure,graphicx}
\usepackage{mathrsfs}
\usepackage{psfrag}
\usepackage{tikz,pgfplots}
\usepackage{booktabs}
\usepackage{url}
\usepackage{bm}
\usepackage{tikz,amsmath,graphicx,xcolor,soul,verbatim}
\usetikzlibrary{arrows}
\usepackage{verbatim}
\usepackage{enumerate}
\usepackage[skip=2pt,font=footnotesize]{caption}

\newtheorem{lemma}{Lemma}
\newtheorem{proposition}{Proposition}
\newtheorem{example}{Example}

\newcommand*{\rom}[1]{\expandafter\@slowromancap\romannumeral #1@}
\newcommand{\be}{\begin{equation}}
\newcommand{\ee}{\end{equation}}

\newcommand{\Xc}{\mathcal{X}}

\newcommand{\Vc}{\mathcal{V}}
\newcommand{\Tc}{\mathcal{T}}

\newcommand{\bfx}{\boldsymbol{x}}

\newcommand{\bfy}{\boldsymbol{y}}
\newcommand{\un}{\underline}
\newcommand{\sfP}{{\sf{P}}}

\newcommand{\syn}{\textsf{syn}}
\newcommand{\summ}{\mathsf{sum}}

\makeatother

\definecolor{mycolor1}{rgb}{0.00000,0.44700,0.74100}%
\definecolor{mycolor2}{rgb}{0.85000,0.32500,0.09800}%
\definecolor{mycolor3}{rgb}{0.92900,0.69400,0.12500}%
\definecolor{mycolor4}{rgb}{0.49400,0.18400,0.55600}%

\allowdisplaybreaks
\IEEEoverridecommandlockouts

\begin{document}

\title{Coding for Deletion Channels with Multiple Traces}

\author{\IEEEauthorblockN{Mahed Abroshan}
\IEEEauthorblockA{University of Cambridge\\
\tt{ma675@cam.ac.uk}}
\and
\IEEEauthorblockN{Ramji Venkataramanan}
\IEEEauthorblockA{University of Cambridge\\
\tt{rv285@cam.ac.uk}}
\and
\IEEEauthorblockN{Lara Dolecek}
\IEEEauthorblockA{ECE Department, UCLA\\
\tt{dolecek@ee.ucla.edu}}
\and
\IEEEauthorblockN{Albert Guill\'{e}n i F\`{a}bregas}
\IEEEauthorblockA{ICREA \& Universitat Pompeu Fabra\\
University of Cambridge\\
\tt{guillen@ieee.org} \vspace{-10pt}}
\thanks{This work was supported in part by the European Research Council under Grant 725411, by the Spanish Ministry of Economy and Competitiveness under Grant TEC2016-78434-C3-1-R,  and by NSF grants CCF-1718389 and CCF-1527130.
}}

\maketitle
\begin{abstract}
Motivated by the sequence reconstruction problem from traces in DNA-based storage, we consider the problem of designing codes for the deletion channel when multiple observations (or traces) are available to the decoder. We propose simple binary and non-binary codes based on Varshamov-Tenengolts (VT) codes. The proposed codes split the codeword in blocks and employ a VT code in each block. The availability of multiple traces helps the decoder to identify deletion-free copies of a block, and to avoid mis-synchronization while decoding. The encoding complexity of the proposed scheme is linear in the codeword length; the decoding complexity is linear in the codeword length, and quadratic in the number of deletions and the number of traces. 
The proposed scheme offers an explicit low-complexity technique for correcting deletions using multiple traces. 
\end{abstract}

\section{Introduction}

We consider the problem of coding for a deletion channel which,  given an input sequence,  produces $t$ output sequences. Each output sequence, known as a trace, is produced by deleting $k$ symbols from the length $n$ input sequence. The deletion pattern for each trace is assumed to be independent, and the locations of the deletions within a trace are uniformly random. (Therefore it is possible that two or more traces are identical.)

The problem of recovering coded information from multiple traces is relevant in DNA-based storage systems \cite{kiah2016codes}.  When  retrieving information by sequencing stored DNA,   each trace may contain errors that are a combination of deletions, insertions, and substitutions. 
In this paper, using the stylized  model of a channel that introduces only deletions, we aim to understand the coding advantage obtained by having multiple traces. In particular, we show how one can use simple codes with  efficient encoding and decoding to achieve small probability of error under suitable assumptions.

The problem of reconstructing a sequence using traces from a deletion channel has been studied in several works, with the goal being either exact recovery of the sequence \cite{lev2001,holenstein2008trace,peresaverage}, or an estimate  \cite{s2018maximum}. In these papers, the sequence  can be an arbitrary one from the underlying alphabet, i.e., it need not originate from a codebook. A few recent works study  the reconstruction of a \emph{coded} sequence under different trace models. The paper  \cite{gabrys} analyzes the minimum number of deletion channel traces required to recover a sequence  drawn from a single-deletion correcting code, and \cite{sala} considers a similar problem for the insertion channel.  The problem of reconstructing a coded sequence from the multiset of its substrings is studied in \cite{2018unique}. From an information-theoretic perspective, \cite{haeupler2014repeated} characterizes  the capacity of the multiple trace i.i.d. deletion channel  as the deletion probability $p \to 0$. 
Recently, coding  for the multiple trace i.i.d. deletion channel  was independently studied in \cite{cheraghchi2019coded}. The construction proposed in \cite{cheraghchi2019coded}  is based on marker codes, and is shown to achieve vanishing error probability with  redundancy $O(n/\log n)$ and $\exp\left(O(\log^{2/3}n)\right)$ traces.

\emph{Overview of the coding scheme}:  Our code construction is based on the family of Varshamov-Tenengolts (VT) codes \cite{VT65,Tenengolts84}. VT codes are single deletion correcting codes, and can be constructed for any finite alphabet and any block length. (VT codes will be reviewed in the next section). Each codeword in our code is a concatenation of blocks, with each block drawn from a predetermined VT codebook.

We illustrate the idea using the following binary example, which shows a codeword  $\bfx$ of length $15$ with three blocks, each  of which is a sequence from a length $5$  binary VT code. The channel produces  two traces, $\bfy_1$ and $\bfy_2$, by deleting the underlined bits:
\begin{align*}
& 1\un{0}001 \hspace{2mm} 110\un{1}1 \hspace{2mm} 01010 \longrightarrow \bfy_1 = 1001110101010 \\
& 10001 \hspace{2mm} 1\un{10}11 \hspace{2mm} 0101\un{0} \longrightarrow \bfy_2 = 100011110101
\end{align*}

The decoder operates in two phases. In phase 1, it identifies blocks that are deletion-free in at least one of the  traces. Each block for which a deletion-free copy is identified in one of the traces is recovered by inserting the required bits in the other traces.  In the example above, block $1$ has no deletions in $\bfy_2$, so $\bfy_2$ is used to correct the first block of $\bfy_1$; similarly,  block $3$ has no deletions in $\bfy_1$.  Assuming that there were no errors, at the end of this phase the decoder has corrected all  blocks which are deletion-free in at least one trace. We call  the remaining blocks  `congested'. In the example, block $2$ is congested as both traces have bits deleted in this block.

In phase 2, the decoder attempts to correct the congested blocks, i.e.,  blocks for which no clean copy was found in phase 1. In the example,  since $\bfy_1$ has one deletion in block $2$, the block can be corrected  using the VT decoder.  Since blocks $1$ and $3$ were corrected in phase 1, $\bfx$ is recovered.   

However,  decoding errors may occur in either phase. In phase $1$, we may wrongly identify a block as deletion-free in a trace, which leads to errors in the starting positions of other blocks. In phase $2$, a congested block (or set of consecutive congested blocks) may not be correctable with the VT code, because of too many deletions in each trace. In phase $1$, wrongly identifying a trace as having a deletion-free copy of a block will lead to an unusually large number of insertions when correcting the other traces using this copy. This can be used to discard accidental matches in phase $1$. We show via numerical simulations that the probability of phase 1 error  decreases  with $n_b$, the length of each block. For the phase 2 error, under the assumption that the locations of the deletions within each trace are uniformly random,  we obtain a bound  that decreases exponentially with the number of traces.

The rate of the  code is equal to  the rate of a VT code of length $n_b$, which is close to $\log q - {\log n_b}/{n_b}$, where $q \geq 2$ is the alphabet size. (The precise values are given in Section \ref{sec:encoding}.) The decoding complexity is $\mathcal{O}(t^2 k^2 n)$. Therefore the proposed scheme offers  an explicit, efficient technique for recovering from deletions using multiple traces. Due to its low-complexity, it can be well suited to a variety of applications, including DNA-based storage.

\emph{Notation}: Sequences are denoted using bold letters, and  scalars with plain font. 
For  $\bfx=x_1x_2\cdots x_n$, the subsequence $x_ix_{i+1}\cdots x_{j}$ is denoted by $\bfx(i:j)$.

\section{Code Construction} \label{sec:encoding}

We first review the family of VT codes, and then present the code construction.

{\em Binary VT codes:}
The VT syndrome of a  binary sequence $\bfx =x_1x_2\cdots x_n$  is defined as 
\begin{equation}
\syn(\bfx)=\sum_{j=1}^n j \,x_j \  \  (\text{mod}\  (n+1)).
\end{equation}
For positive integers $n$ and $0\leq s\leq n$, we define the VT code of length $n$ and syndrome $s$, denoted by 
\be
\Vc\Tc_s(n) = \big\{\bfx \in \{0,1\}^n: \syn(\bfx)=s\big\} ,
\ee
as the set of sequences $x$ of length $n$ for which $\syn(\bfx)=s$.  The $(n+1)$ sets $\Vc\Tc_s(n) \subset \{0,1\}^n$, for $0\leq s\leq n$, partition the set of all binary sequences of length $n$.  Each of these sets $\Vc\Tc_s(n)$ is a zero-error single-deletion correcting code. The VT encoding and decoding complexity is linear in the code length $n$ \cite{Sloane00,AbdelFer98}.

{\em Non-binary VT codes:}
VT codes were extended to non-binary alphabets  in \cite{Tenengolts84}. 
Let the alphabet be $\Xc = \{0, \ldots, q-1 \}$, with $q >2$.   
For each sequence $\bfx\in \Xc^n$, define a length $(n-1)$ auxiliary binary sequence 
$\boldsymbol{\alpha}(\bfx) = \alpha_2,\dotsc,\alpha_n$  as follows. For $2 \leq j \leq n$,
\be \alpha_j =
  \begin{cases}
    1  & \quad \text{if } x_j \geq x_{j-1}\\
    0  & \quad \text{if } x_j< x_{j-1} \\
  \end{cases}
\ee
We also define the modular sum as
\be
\summ(\bfx)=\sum_{j=1}^n x_j \quad (\text{mod } q).
\ee
A $q$-ary VT code with length $n$ and parameters $(a,c)$ is defined as \cite{Tenengolts84}
\be
\Vc\Tc_{a,c}(n) = \big\{\bfx \in \Xc^n: \syn(\boldsymbol{\alpha}(\bfx) )=a,\ \summ(\bfx)=c\big\},
\label{eq:VT_nonbin}
\ee
for $0\leq a\leq n-1$ and $c\in \Xc$. Similarly to the binary case,  the sets $\Vc\Tc_{a,c}(n)$  partition the space $\Xc^n$ of all $q$-ary sequences of length $n$ into $qn$ sets. Each set is a single deletion correcting code.  The complexity of the encoding and decoding $q$-ary VT codes is linear in the code length $n$ \cite{Tenengolts84,abroshan2018efficient}.

\subsection{Code construction}

Codewords of length $n$ are constructed by concatenating $l$ blocks of VT codewords from the relevant alphabet. Each  block has length $n_b$ (therefore $n=ln_b$). 

{\em Binary code:}  Each block $i$ ($1 \leq i \leq l$) is a binary VT codeword with a predetermined  VT syndrome $a_i$, known to both the encoder and the decoder.  To encode each block, one can use the systematic VT encoder in \cite{AbdelFer98} that maps $n_b-\lceil \log (n_b+1) \rceil$ bits to a length $n_b$ VT sequence with the desired syndrome.  The rate of the code will be
\be R=1- \lceil \log (n_b+1) \rceil /n_b.  \label{eq:bin_rate} \ee

{\em Non-binary code:} The code construction  is very similar to the binary case. Each block  is encoded separately, and belongs to a known non-binary VT class, as defined in \eqref{eq:VT_nonbin}. 
There are $qn_b$ non-binary VT classes, so there exists a class with at least $\frac{q^{n_b}}{qn_b}$ sequences. Using this class for  encoding each block  induces the following lower bound for the rate of the code:
\be 
R \geq \log q - (\log n_b + \log q)/n_b. 
\label{eq:q_rate} 
\ee

The VT syndrome/class for each of the $l$ blocks can be arbitrarily chosen. Choosing the largest of the VT classes of length $n_b$ sequences maximises the rate of the code.

\section{Decoding}\label{sec:dec}
The goal of the decoder is to reconstruct $\bfx$ using the  traces $\bfy_1,\bfy_2, \cdots, \bfy_t$, each   obtained by deleting  $k$ symbols from $\bfx$.  We describe the binary decoder, and then outline the main differences for the non-binary case. 
We  explain the main ideas using examples, and then specify the decoding algorithm in detail.
We remark that the same decoding algorithm can be applied when the number of deletions in one or more traces is less than $k$.

\subsection{Phase 1} \label{subsec:dec_phase1}
Consider  block $i$ of the codeword, for $1 \leq i \leq l$.
If  the starting position of block $i$ within each trace is known, then the decoder can compute the VT syndrome of the  length $n_b$ sequence from the starting position,  for each trace. If there is a trace for  which the computed syndrome matches with $a_i$ (the correct syndrome for block $i$), then the trace can be used to correct this block within other traces.   The following example illustrates this idea.

\begin{example}
Consider a codeword with $l=3$ blocks,  with each block of length $n_b=5$. Thus $n=15$.  Let
\begin{align*} \bfx= 01001\hspace{2mm} 11 00 1\hspace{2mm} 11111 \end{align*}
be  the transmitted codeword. The VT syndromes of the blocks are $a_1=1$, $a_2=2$, and $a_3=3$. Suppose that the decoder receives two traces, each with $k=2$ deletions. The underlined bits are deleted from $\bfx$ to produce $\bfy_1$ and $\bfy_2$:
\begin{align*} 
& 01\underline{00}1 \hspace{2mm} 11001 \hspace{2mm} 11111 \longrightarrow \bfy_1= 011 11001  11111 \\
& 01001 \hspace{2mm} \underline{1}1 00 \underline{1} \hspace{2mm} 11111\longrightarrow \bfy_2= 01001 100 11111
\end{align*}
The decoder first computes the VT syndromes of $\bfy_1(1:5)$ and $\bfy_2(1:5)$. We have  
$$\syn(\bfy_1(1:5))=2 \  \text{ and } \  \syn(\bfy_2(1:5))=1.$$ Since $a_1=1$, the decoder assumes that $\bfy_2(1:5)$ is the first block of $\bfx$, and uses it  to correct the first block of $\bfy_1$ by inserting the two missing bits. The decoder then considers the second block, whose starting position is now known for each trace. Finding that $\syn(\bfy_1(4:8))=a_2=2$, it assumes this sequence is the second block of $\bfx$, and uses it  to correct $\bfy_2$ by inserting two bits. Since there are no deletions in the third block, the decoder finds   $\syn(\bfy_1(9:13)) = \syn(\bfy_2(9:13))= a_3=3$, and  stops.
\end{example}

A decoding error may occur if the VT syndrome of a  block in a trace accidentally matches the correct value. In this case, an incorrect sequence will be used to correct the block in all other traces, potentially introducing multiple errors. The following example shows that how other traces can help to identify and discard such accidental matches.

\begin{example}
\label{ex:acc1}
Consider a codeword with $l=2$ blocks, with each block of length $n_b=10$. The VT syndromes of the blocks are $a_1=a_2=5$. Let the transmitted codeword be
\begin{align*}  \bfx=1000111100\hspace{2mm} 0011101100. \end{align*}
 There are two traces, each with $k=2$ deletions (underlined bits are deleted):
\begin{align*} 
& 100\underline{0}111\underline{1}00\hspace{2mm} 0011011100 \longrightarrow \bfy_1=10011100 0011011100 \\
& \underline{1}000111100 \hspace{2mm} 001\underline{1}011100 \longrightarrow \bfy_2=000111100 001011100
\end{align*}
The decoder finds that $\syn(\bfy_1(1:10))=5$ (this is an instance of an accidental match), and $\syn(\bfy_2(1:10))=0$. It assumes that $\bfy_1(1:10)$ is the correct block, and uses it to fix $\bfy_2$.
It does this by comparing $\bfy_1(1:10)$ with $\bfy_2$, and inserting the required bits to get $\hat \bfy_2(1:10)=\underline{1}00\underline{111}0\underline{000}$. Since there are two deletions in $\bfy_2$,  exactly two inserted bits are required to recover the codeword. However, since $7$ bits need to be inserted into $\bfy_2$ to get $\bfy_1(1:10)$ and $k=2$, the decoder realizes that $\bfy_1(1:10)$ is an accidental match.  
\end{example}
The above example shows that when an accidentally matched block is used as the model to correct other traces, the number of inserted bits is likely to be large. Hence the decoder can distinguish between an accidental match and a correct match  in most cases.

\emph{Congested blocks and resynchronization}. There may be  blocks that have undergone at least one deletion in each of the traces. These blocks are called \emph{congested},   as a correct match for them cannot be found in any of the traces. In Example \ref{ex:acc1}, the first block is congested  as there are deleted bits in both of the traces. As the decoder proceeds from left to right in phase 1, it needs to resynchronize whenever it identifies a congested block. It does so by testing all possible starting positions for the next block in each trace.  

Assume block $i$ is congested, and consider a trace for which that the decoder has inferred that there are $d <k$ deletions up to block $(i-1)$. The decoder needs to test $(k-d)$ possible starting positions for  block $(i+1)$ in this trace. It computes  the VT syndromes of the length $n_b$ sequences starting from each of these positions,  and checks for a match with the correct syndrome $a_{i+1}$. If a match is found,  it is used to correct the other traces.  It repeats this process for each trace, testing all possible starting positions for block $(i+1)$, and checking whether a match is found for the correct VT syndrome.   If the decoder finds  one or more syndrome matches among those tested, it  chooses the one that requires the minimum number of insertions (across all traces) for  correcting  block $(i+1)$. 
 
 When block $i$ is identified as congested, it is  possible that block  $(i+1)$ is also congested (i.e., has deletions in all the traces). In this case, no matches may be found among all the tested starting positions for block $(i+1)$. The decoder then tries to synchronize  by testing all possible starting positions for block $(i+2)$. 
 
 \subsection{Phase 2} \label{subsec:dec_phase2}
 
 At the end of phase $1$, if there are no errors, the decoder has corrected all blocks for which there is at least one trace with a deletion-free copy of the block. Each remaining block is congested, and is either: i) isolated, i.e., the bits corresponding to the block in each trace are known, or ii) part of an isolated  set of consecutive congested blocks.

  In the second phase, the decoder uses the VT syndromes to correct as many congested blocks as possible.  For each congested set of $r$ consecutive  blocks ($r \geq 1$, with $r=1$ corresponding to a single congested block), the decoder  can infer the number of deleted bits within each trace. It uses this information, and attempts to correct the congested blocks as follows. For a congested set of $r$ consecutive  blocks ($r \geq 1$), the decoder looks for a trace with exactly $r$ deletions. If such a trace exists, then this set of blocks can be corrected using that trace and the known VT syndromes of the $r$ blocks. On the other hand, a congested set of $r$ consecutive blocks cannot be corrected if it has  at least $(r+1)$ deletions in each trace.

\begin{example}
Consider a codeword with $l=4$ blocks, each of length $n_b=5$. The VT syndromes  are $a_1=a_2=a_3=a_4= 0$, and 
\begin{align*}
\bfx  &= 11100 \ 10001 \ 10001 \ 01010.
\end{align*}
There are two traces, with $4$ deletions in the first and $3$ in the second:
\begin{align*}
1110\underline{0} \ 1\underline{0}001 \ 10001 \ \underline{0}101\underline{0} \to \bfy_1&= 1110 1001 10001  101\\
11100 \ 100\underline{0}1 \ 10001 \ 0\underline{1}0\underline{1}0\to 
\bfy_2&= 11100 1001  10001  000.
\end{align*}
The first block is recovered using $\bfy_2$, using which the block is corrected in $\bfy_1$. The second block is congested, and neither trace provides a match for its VT syndrome. The decoder therefore tests the possible starting positions for the third block. Consider the first trace, which has a total of $4$ deletions. Since there was one deletion in the first block, there are three  possible starting positions for the third block: bits $7,8$ and $9$ of $\bfy_1$. Similarly,  bits $8,9$ and $10$  of $\bfy_2$ are the possible starting positions for the third block. 

The decoder therefore computes the VT syndrome of $\bfy_1(9:13), \bfy_1(8:12), \bfy_1(7:12)$, and $\bfy_2(10:14)$, $\bfy_2(9:13)$, 
$\bfy_2(8:12)$. Among these, the only one that satisfies the correct syndrome $a_3=0$ is $\bfy_1(9:13) = \bfy_2(10:14)= 10001$. This indicates that there is one deletion in the second block, in each of the traces. Thus the second block can be recovered using the VT decoder in phase 2. With the first three blocks synchronized, the decoder attempts to correct the fourth. The fourth block has  two deletions in both traces.  As the VT decoder can only correct a single deletion, the decoder declares an error due to an unresolvable congestion.
\end{example}

 In the next section, we derive a bound (Propositon \ref{prop:phase2_err}) on the probability of  phase 2 error, caused by an unresolvable congestion like the one above.

\subsection{Decoding algorithm (for binary alphabet)}\label{sec:dec:alg}
We now describe the decoder in detail.  
Denote the number of deletions in the $j$th trace by  $k_j$, recalling that $k_j\leq k$ for $1\leq j \leq t$. 
 
 {\em \underline{Phase 1}}

\un{Block 1}: Compute the VT syndrome of $\bfy_j(1:n_b)$, for $1\leq j\leq t$. If the computed syndrome for trace $j$  is equal to $a_1$,  consider  $\bfy_j(1:n_b)$ as a candidate for the first block of the codeword, and use it to correct the other traces.  In the process, if the total number of bits inserted into any trace exceeds the number of deletions in it, discard $\bfy_j$ from the list of candidates. If the final list of candidates is non-empty, pick  one that leads to the fewest total insertions in the other traces. 
If the final list  of candidates is empty, declare block $1$ congested and proceed to the second block.

 \un{Block $i>1$}: There are two possibilities:
\begin{enumerate}
\item If block $(i-1)$ is not congested: The starting position of  the $i$th block is known in each trace. As in block 1,  for each trace compute the VT syndrome for the length $n_b$ sequence from the starting position, and compare  with $a_i$. Each  sequence whose VT syndrome matches $a_i$ is a candidate. Use each candidate sequence to correct the other traces; if the total number of bits inserted in any trace (up to this point in decoding) exceeds the number of deletions in it, discard the sequence from the list of candidates. If the final list of candidates is non-empty, pick one that leads to the fewest total insertions in the other traces.  If the final list  list of candidates is empty, declare block $i$ congested, and proceed to the next block.

\item If block $(i-1)$ is congested: The starting position of block $i$ is not known. Suppose that blocks $(i-1)$ to $(i-c)$ are congested (where $c \geq 1$). Since block $(i-c-1)$ is not congested, for each trace the decoder can infer the total number of deletions up to block $(i-c-1)$. Denote this number  by $d_j$ for trace $j$. Then the starting position of  the block $i$ in trace $j$ is a number between $(i-1)n_b-c-d_j+1$ and $(i-1)n_b-k_j+1$, where $k_j$ is the total number of deletions in trace $j$. Compute the VT syndrome for each of these $(k_j- c - d_j+1)$ possibilities, and compare  with $a_i$. If there is a sequence whose syndrome matches, add it to the list of candidates and correct  the other traces using this sequence. Since the starting position of block $i$ is not known, when correcting using a candidate sequence, we need to consider all the possible starting positions of block $i$ in the other traces. Pick the starting position that  results in the minimum number of inserted bits. (If there is more than one starting position that gives the minimum, we pick the rightmost one.) As before, discard a candidate if the number of bits inserted in trace $j$ is larger than $k_j-c-d_j$ for some $j$. 

 If the final list of candidates is non-empty, pick one that leads to the fewest total insertions in the other traces. This process also  gives the starting positions for block $(i+1)$ in each trace.  If the final   list of candidates is empty, declare block $i$ congested, and proceed to the next block.

\end{enumerate}

 {\em \underline{Phase 2}}
 
 Consider each  congested set of $r$ consecutive blocks  separately,  for $1 \leq r \leq k$. For each of these congested sets,   the decoder knows the number of deletions in each trace.  For a congested set with $r$ blocks,  if each trace has more than  $r$ deletions in the congested set, the decoder declares an error. Otherwise the decoder finds a trace with exactly $r$ deletions in the congested set, i.e., exactly one deletion per block. The decoder corrects these blocks using the VT decoder, and uses them to correct the other traces by inserting the appropriate bits.  During this process, if  the number of inserted bits does not match the number of deletions in the trace within the congested set, the decoder declares an error.

\subsection{Non-binary alphabet}\label{sec:non-binary}
The decoding is similar to the binary case. The only difference is that the VT syndrome of a non-binary sequence is a pair of numbers. Therefore, when we comparing VT syndromes of two sequences in the first phase, both numbers in the pair should be compared.  In the second phase,  the decoder uses the  non-binary VT decoder from \cite{Tenengolts84} to recover a block with a single deletion.

\subsection{Decoding complexity}
In phase 1, for a block for which the starting position is unknown, the decoder computes at most $k$  VT syndromes of  length $n_b$ sequences in each of the $t$ traces.  
For each matched syndrome, the decoder needs to check inserted bits in at most $k$ blocks in the other $(t-1)$ traces. Since there are $l$ blocks, and 
$n=n_b l$, the complexity for the first phase is $\mathcal{O}(t^2k^2 n)$.

In  phase 2, the VT decoder is used in at most $l$ blocks (each of length $n_b$), and then uses the recovered sequence to correct the block in the other traces. Since the VT decoder has linear complexity,  the complexity for phase 2 is  $\mathcal{O}(tn)$.

\section{Error probability and Simulation results}\label{sec:error}
\subsection{Phase 1 errors}\label{sec:ph1:error}
In the first phase of decoding, an error can occur in two ways. 
 First, an accidental match may lead to a block being wrongly identified as deletion-free in a trace; this is then used to correct the block in other 
traces. Second, when a congested block is identified, the decoder may pick a wrong starting position for the next block. 
As shown in Example \ref{ex:acc1}, an accidental match in a trace can be often detected by the decoder when it  leads to a large number of inserted bits in the other traces. This detection feature  makes it hard  to derive a rigorous bound for the phase 1 error.

Without the detection feature,  the probability of an accidental VT match in the binary case will be inversely proportional to $n_b$, the length of the block. Indeed,  the family of $(n_b+1)$ VT codes partitions the space of length-$n_b$  binary sequences into approximately equal-sized sets of size $\sim 2^{n_b}/(n_b+1)$. Hence the probability that a binary sequence picked uniformly at random will match a given VT syndrome is close to $\frac{1}{(n_b+1)}$.  

\subsection{Phase 2 errors}
Errors in  the second phase of the decoding are due to unresolvable congestion. Recall that unresolvable congestion occurs if, for some $1 \leq r \leq k$,  there is a set of $r$ consecutive congested blocks with at least $(r+1)$ deletions in each  trace. The following proposition bounds the probability of phase 2 error, denoted by 
$\sfP_{e_2}$.

\begin{proposition}
\label{prop:phase2_err}
Consider a code with $l$ blocks, and a channel that introduces at most $k$ deletions in each of the $t$ traces. If  $k<l$ and the locations of deletions within each trace are uniformly random,  the probability of phase 2 error satisfies
\begin{align}
& \sfP_{e_2}  \nonumber  \\
%
%
&\leq l \Big((1-p_0-p_1)^t \hspace{-1pt} + \left((1-p_0)^2-p_1p_1'\right)^t \hspace{-1pt} + 
\frac{(1-p_0)^{3t}}{1- (1-p_0)^t}\Big) \nonumber
\end{align}
where, for $s = 0,1$,
\be 
\label{eq:p:i}
p_s= \frac{\binom{(k-s)+(l-1)-1}{k-s}}{\binom{k+l-1}{k}}, \ \text{ and } \ 
p'_1=\frac{\binom{(k-2)+(l-2)-1}{k-2}}{\binom{(k-1)+(l-1)-1}{k-1}}.
\ee
\end{proposition}

We note that the phase 2 error probability (and the bound) depends only on $k$ and $l$, and on neither $n_b$ nor the alphabet size.  The restriction $k < l$ is natural, as otherwise the expected number of deletions per block  would be greater than $1$.

\begin{proposition}
\label{prop:phase2_err}
Consider a code with $l$ blocks, and a channel that introduces at most $k$ deletions in each of the $t$ traces. If  $k<l$ and the locations of deletions within each trace are uniformly random,  the probability of phase 2 error satisfies
\begin{align}
& \sfP_{e_2}  \nonumber  \\
&\leq l \Big((1-p_0-p_1)^t+ \left((1-p_0)^2-p_1p'_1\right)^t+\sum_{r=3}^{k-1} (1-p_0)^{rt} \Big) \label{eq:pe2_bnd1}\\
&\leq l \Big((1-p_0-p_1)^t \hspace{-1pt} + \left((1-p_0)^2-p_1p_1'\right)^t \hspace{-1pt} + 
\frac{(1-p_0)^{3t}}{1- (1-p_0)^t}\Big)
\end{align}
where, for $s = 0,1$,
\be 
\label{eq:p:i}
p_s= \frac{\binom{(k-s)+(l-1)-1}{k-s}}{\binom{k+l-1}{k}},
\ee
and 
\be 
p'_1=\frac{\binom{(k-2)+(l-2)-1}{k-2}}{\binom{(k-1)+(l-1)-1}{k-1}}.
\label{eq:p1pr}
\ee
\end{proposition}

We note that the phase 2 error probability (and the bound) depends only on $k$ and $l$. It does not depend on either $n_b$ or the alphabet size.  The restriction $k < l$ is natural, since otherwise the expected number of deletions per block  would be greater than $1$.

\begin{IEEEproof} 
For $1 \leq i \leq l$ and $1 \leq r \leq k$, 
let $Z_{i,r}$ be an indicator random variable with $Z_{i,r} =1$ if the $i$th block is in an unresolvable congestion of exactly $r$ consecutive blocks, and  $Z_{i,r} =0$ otherwise. Let
\be 
Z= \sum_{r=1}^{k-1} \sum_{i=1}^l \frac{1}{r} Z_{i,r}.
\label{eq:Zdef}
\ee
For each $r$, the inner sum in \eqref{eq:Zdef} counts the number of distinct sets of $r$ consecutive congested blocks. Therefore  $Z$ is the total  number of  distinct congested {sets}, where a congested set is a set of of $r$ consecutive congested blocks, for some $r \geq 1$.
Hence $\sfP_{e_2}=\mathbb{P}(Z\geq 1)$. Using Markov's inequality,
\begin{align}
  \label{Markov}
        \mathbb{P}(Z\geq 1)&\leq \mathbb{E}[Z]
=  \sum_{i=1}^l \sum_{r=1}^{k-1} \frac{1}{r} \mathbb{E} [Z_{i,r}]
\end{align}

The probability that a given block has exactly $s$ deletions (for $0 \leq s \leq k$) is given by $p_s$ in 
\eqref{eq:p:i}. Indeed, since the locations of the $k$ deletions are uniformly random, the probability of a block having $s$ deletions is the proportion of non-negative integer solutions of $x_1+x_2+\cdots +x_l=k$ with  $x_1=s$.

A block can be in an unresolvable congestion only if it has more than one deletion in each of the traces. Therefore,
\begin{equation}
\mathbb{E}[Z_{i,1}]\leq (1-p_0-p_1)^t.    \label{eq:X1}
\end{equation}
To find an upper bound for $\mathbb{E}[Z_{i,r}]$ for $r\geq 2$, we  need the following lemma. 

\begin{lemma} For a given trace and $r$ blocks ($1\leq r\leq k$), denote by $q_r$ the probability of at least one deletion occurring in each of the $r$ blocks. Then 
\be 
q_r \leq (1- p_0)^r.
\label{eq:qr_eqn}
\ee 
\end{lemma}
\begin{IEEEproof}
 We prove this by using induction on $r$. For $r=1$, we have $q_1 =(1-p_0)$. Now assume that \eqref{eq:qr_eqn} holds for  $q_u$, for some $u<r$. For $s \geq u$, we  write $q_u(s)$ for the probability of $s$ deletions occurring in a given set of $u$ consecutive blocks, with at least one deletion in each of them. Clearly,  $q_u = \sum_{s=u}^k q_u(s)$. We then have 
\begin{align}
    q_{u+1}&=\sum_{s=u}^k q_u(s) \left(1- \frac{\binom{(k-s)+(l-u)-2}{k-s}}{\binom{(k-s)+(l-u-1)}{k-s}} \right)\\
    &=\sum_{s=u}^k q_u(s) \left(1- \frac{l-u-1}{(k-s)+(l-u-1)} \right)\\
    &\leq \sum_{s=u}^k q_u(s) \left(1- \frac{l-1}{k+l-1} \right) \label{eq:rslk_bnd} \\
    &=\sum_{s=u}^k q_u(s) \left(1- p_0 \right)\\
    &\leq (1-p_0)^{u+1}. \label{eq:final_eq}
\end{align}
In the chain above, it can be  verified that \eqref{eq:rslk_bnd} holds when $l >k$ and $s \geq r$. Eq. \eqref{eq:final_eq} is obtained using the induction hypothesis: $ \sum_{s=u}^k q_u(s) = q_u \leq (1-p_0)^u$.
\end{IEEEproof}
Using the lemma, the probability that two consecutive blocks, say $i$ and $(i+1)$, have at least three deletions in each trace (and are hence unresolvable)  is bounded by 
$[(1-p_0)^2-p_1p'_1]^t$. Here $p_1'$ defined in \eqref{eq:p1pr} is the probability of  block $(i+1)$ having one deletion \emph{given} that block $i$ has one deletion. Therefore, considering the event of unresolvable congestion  either in the pair of blocks $\{ (i-1), i \}$ or in blocks $\{ i, (i+1) \}$, we have 
\be
\label{eq:X2}
    \mathbb{E}[Z_{i,2}] \leq 2\left((1-p_0)^2-p_1p'_1\right)^t.
\ee
For $r > 2$,  consider a set of $r$ consecutive blocks, say $i, \ldots, (i+r-1)$. The probability of congestion in this set of blocks is $q_r^t$, which by \eqref{eq:qr_eqn} is bounded by $(1-p_0)^{rt}$. Hence,
\begin{align}
    \mathbb{E}[Z_{i,r}]&\leq r (1-p_0)^{tr}, \label{eq:Xr}
\end{align}
where we use the fact that a given block $i$ is part of (up to) $r$ different sets of $r$ consecutive blocks. Using \eqref{eq:X1}, \eqref{eq:X2}, and \eqref{eq:Xr} in \eqref{Markov}  yields the result of the proposition.
\end{IEEEproof}

\begin{figure}[t]
\centering
\begin{tikzpicture}
\begin{axis}[%
width=0.8\columnwidth,
at={(0in,0in)},
scale only axis,
xlabel={Length of block $n_b$},
ylabel={Probability of error},
xmin=20,
xmax=100,
ymin=0,
ymax=0.0012,
axis background/.style={fill=white},
legend style={legend cell align=left, align=left, draw=white!15!black}
]
\addplot [color=mycolor1, mark=asterisk, mark options={solid, mycolor1}]
  table[row sep=crcr]{%
20	0.001022\\
30	0.0005579\\
40	0.0003821\\
50	0.000293\\
60	0.0002391\\
70	0.0002028\\
80	0.000184\\
90	0.0001598\\
100	0.0001516\\
};
\addlegendentry{$q=2$}

\addplot [color=mycolor2, mark=o, mark options={solid, mycolor2}]
  table[row sep=crcr]{%
20	0.0002103\\
30	0.0001376\\
40	0.0001055\\
50	8.96e-05\\
60	9.22e-05\\
70	8.45e-05\\
80	8.41e-05\\
90	8.09e-05\\
100	8.24e-05\\
};
\addlegendentry{$q=4$}

\addplot [color=mycolor3, dashed]
  table[row sep=crcr]{%
20	7.63e-05\\
30	7.63e-05\\
40	7.63e-05\\
50	7.63e-05\\
60	7.63e-05\\
70	7.63e-05\\
80	7.63e-05\\
90	7.63e-05\\
100	7.63e-05\\
};
\addlegendentry{$\sfP_{e_2}$ simulation}
\end{axis}
\end{tikzpicture}
\caption{Probability of error for different block lengths when $l=7,k=4$, and $t=5$.}
\vspace{-5pt}
\label{plot:n}
\end{figure}
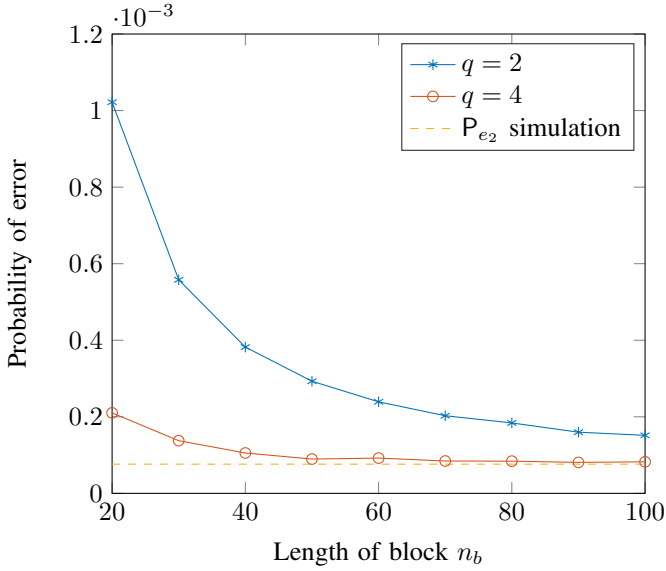

 
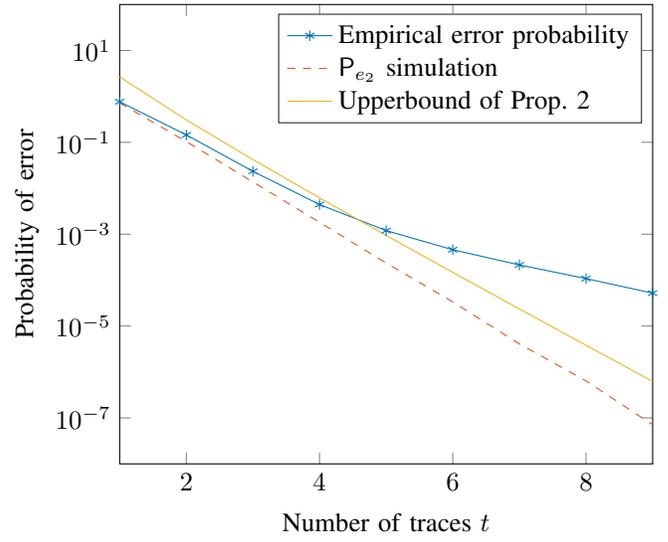
\begin{figure}[t]
\centering
\begin{tikzpicture}
\begin{axis}[%
width=0.8\columnwidth,
at={(0.758in,0.481in)},
scale only axis,
xlabel={Number of traces $t$},
ylabel={Probability of error},
xmin=1,
xmax=9,
ymode=log,
ymin=1e-08,
ymax=100,
yminorticks=true,
axis background/.style={fill=white},
legend style={legend cell align=left, align=left, draw=white!15!black}
]
\addplot [color=mycolor1, mark=asterisk, mark options={solid, mycolor1}]
  table[row sep=crcr]{%
1	0.7644926\\
2	0.1448775\\
3	0.0233525\\
4	0.0044237\\
5	0.0011961\\
6	0.0004595\\
7	0.0002143\\
8	0.000108\\
9	5.21e-05\\
};
\addlegendentry{Empirical  error probability}

\addplot [color=mycolor2, dashed]
  table[row sep=crcr]{%
1	0.7221872\\
2	0.1039427\\
3	0.0136987\\
4	0.0018204\\
5	0.000236\\
6	3.36e-05\\
7	4.0800e-06\\
8	6.4000e-07\\
9	7.100e-08\\
};
\addlegendentry{$\sfP_{e_2}$ simulation}

\addplot [color=mycolor3]
  table[row sep=crcr]{%
1	2.65714285714286\\
2	0.308072820771233\\
3	0.0421281372201441\\
4	0.00617033816139136\\
5	0.000941680149513452\\
6	0.000147703550473664\\
7	2.36056273822832e-05\\
8	3.8215287506825e-06\\
9	6.24185326959971e-07\\
};
\addlegendentry{Upperbound of Prop. \ref{prop:phase2_err} }
\end{axis}
\end{tikzpicture}%
\caption{Probability of error for a binary code for different values of $t$ when $l=6,k=4$, and $n_b=30$. The code length $n=180$, and the rate is $5/6$. }\label{plot:t}
\vspace{-3pt}
\end{figure}


\begin{figure}[h!]
\centering
\begin{tikzpicture}
\begin{axis}[%
width=0.8\columnwidth,
at={(0in,0in)},
scale only axis,
xlabel={Number of deletions $k$},
ylabel={Probability of error},
xmin=2,
xmax=10,
ymode=log,
ymin=1e-10,
ymax=1,
yminorticks=true,
axis background/.style={fill=white},
legend style={at={(0.35,0.17)},anchor=west,legend cell align=left, align=left, draw=white!15!black}
]
\addplot [color=mycolor1, mark=asterisk, mark options={solid, mycolor1}]
  table[row sep=crcr]{%
2	-0\\
4	2.25e-05\\
6	0.000622\\
8	0.0053468\\
10	0.0276634\\
};
\addlegendentry{$q=2$}


\addplot [color=mycolor3]
  table[row sep=crcr]{%
2	4.70108248747093e-10\\
4	2.15371613102059e-06\\
6	9.2616400953084e-05\\
8	0.000892433869574162\\
10	0.00421406027548268\\
};
\addlegendentry{Upperbound of Prop. \ref{prop:phase2_err} }

\addplot [color=mycolor4, dashed]
  table[row sep=crcr]{%
2	0\\
4	2e-07\\
6	2.32e-05\\
8	0.0004246\\
10	0.0033767\\
};
\addlegendentry{$P_{e_2}$ simulation}
\end{axis}
\end{tikzpicture}%
\caption{Probability of error of a binary  rate $5/6$ code for different values of $k$, the number of deletions. Code parameters are $n_b=30$, $l=10$, and the number of traces $t=5$.}
\label{plot:k}
\vspace{-5pt}
\end{figure}
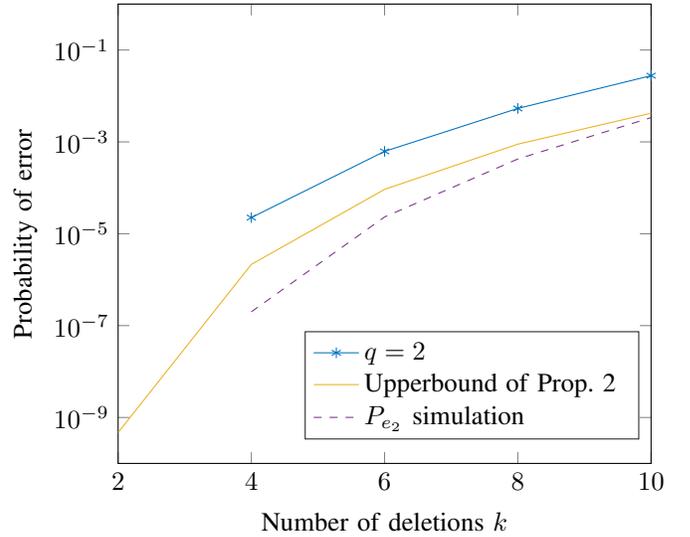

\subsection{Numerical Simulations}
Figure \ref{plot:n} shows the empirical error probability of the code for different values of $n_b$, for $q=2$ (binary) and for $q=4$.  Each codeword consists of $l=7$ blocks, each of length $n_b$.   There are $t=5$ traces, each with $k=4$ deletions at uniformly random locations. 
We note that both the code length  and the rate (cf. \eqref{eq:bin_rate}, \eqref{eq:q_rate}) increase with $n_b$.

Figure \ref{plot:n}  also shows the empirical phase 2 error (dashed line), which does not depend on either $n_b$ or the alphabet.
 Proposition \ref{prop:phase2_err} gives an upper bound of $3.41 \times 10^{-4}$ for the phase 2 error, while the empirical value is $7.63 \times 10^{-5}$.   The difference between the overall and the phase 2 error probabilities can be (roughly) interpreted as the phase 1 error probability. The phase $1$ error caused by wrong matches of VT syndrome decreases with $n_b$, as explained in Sec. \ref{sec:ph1:error}.  Furthermore, we observe that the phase 1 error is smaller (and decreases faster with $n_b$) for $q=4$ than for $q=2$. There are two reasons for this. First, the number of potential VT syndromes for $q >2$ is $qn$, in contrast to the binary case where there are $(n+1)$  VT syndromes. Thus the probability of an accidental match is smaller for the non-binary code. Second, as $q$ increases we expect  an accidental match  to produce more insertions in the other traces, making it is less likely to be accepted as the correct block. This is because  matching of two symbols is less likely in a larger alphabet. 

Figure \ref{plot:t} shows how the error probability decreases with the number of traces  $t$, for a  rate $\frac{5}{6}$ binary code with code parameters held fixed. Each trace has $4$ deletions. As shown in Proposition \ref{prop:phase2_err}, the phase 2 error decays exponentially with $t$. The overall error probability decays  more slowly. Hence for larger values of $t$, phase 1 error becomes the dominant contribution to the overall error probability.

Figure \ref{plot:k} shows the probability of error for different $k$ for a rate $\frac{5}{6}$ binary code with $t=5$ traces. No errors were observed for $k=2$.

\section{Discussion and Future work}

The coding scheme demonstrates that single-deletion correcting codes can be effective in correcting multiple deletions when several traces are available. Having multiple traces helps the decoder in two ways: to identify deletion-free copies of a block, and to avoid mis-synchronization. 

The error performance of the coding scheme can be improved by using more complex decoding in either phase. 
In phase 1, when there are multiple syndrome matches for a block, the decoder could run the decoding procedure for a fixed number of these matches in parallel, using  the ``fewest insertions" rule to decide which candidates to retain. In phase 2, when unresolvable congestion occurs, the decoder could guess a subset of the bits in the block, use the VT decoder to recover the rest, and choose a guess that produces the correct number of insertions in other traces.
Investigating  these improved decoders is part of  ongoing work.   Another idea  for future work is to replace the VT code  in each block with a lower-rate  code capable of correcting more  than one deletion.

\bibliographystyle{ieeetr}
\bibliography{ref}

\end{document}